\documentclass[11pt]{article}

\usepackage[preprint]{acl}

\usepackage{times}
\usepackage{latexsym}

\usepackage[T1]{fontenc}

\usepackage[utf8]{inputenc}

\usepackage{microtype}

\usepackage{inconsolata}

\usepackage[ruled,vlined,linesnumbered]{algorithm2e}
\usepackage{graphicx}
\usepackage{amsmath, amsfonts, amssymb}
 \usepackage{amsmath} 
\usepackage{graphicx}         
\usepackage{booktabs}         
\usepackage{multirow}         
\usepackage{makecell}         
\usepackage{xcolor}   
\usepackage{pifont}   
\usepackage[table]{xcolor}    
\newcommand\myfootnotestyle[1]{\ifcase#1 \or \ding{182}\or \ding{183}\or
\ding{184}\or \ding{185}\or \ding{186}\or \ding{187}%
\or \ding{188}\or \ding{189}\or \ding{190}\or \ding{191}\else *\fi\relax}
\newcolumntype{Y}{>{\centering\arraybackslash}X}

\newcommand{\eg}{\textit{e}.\textit{g}.} 
\newcommand{\Tref}[1]{Tab.~\ref{#1}}

\newcommand{\Fref}[1]{Fig.~\ref{#1}}
\newcommand{\Sref}[1]{Sec.~\ref{#1}}
\newcommand{\Aref}[1]{Alg.~\ref{#1}}
\newcommand{\Asref}[1]{App.~\ref{#1}}

\newcommand{\tool}{\emph{AgentVisor}}
\usepackage{fontawesome}

\title{AgentVisor: Defending LLM Agents Against Prompt Injection via Semantic Virtualization}

\author{
  Zonghao Ying$^{1}$\thanks{Equal contribution.}, 
  Haozheng Wang$^{1}$\footnotemark[1], 
  Jiangfan Liu$^{1}$, 
  Quanchen Zou$^{2}$, \\
  \textbf{Aishan Liu}$^{1}$, 
  \textbf{Jian Yang}$^{1}$, 
  \textbf{Yaodong Yang}$^{3}$, 
  \textbf{Xianglong Liu}$^{1}$ \\
  \\
  $^1$Beihang University \quad $^2$360 AI Security Lab \quad $^3$Peking University \\
  \texttt{}
}

\begin{document}
\maketitle
\begin{abstract}
Large Language Model (LLM) agents are increasingly used to automate complex workflows, but integrating untrusted external data with privileged execution exposes them to severe security risks, particularly direct and indirect prompt injection. Existing defenses face significant challenges in balancing security with utility, often encountering a trade-off where rigorous protection leads to over-defense, or where subtle indirect injections bypass detection. Drawing inspiration from operating system virtualization, we propose \tool{}, a novel defense framework that enforces semantic privilege separation. \tool{} treats the target agent as an untrusted guest and intercepts tool calls via a trusted semantic visor. Central to our approach is a rigorous audit protocol grounded in classic OS security primitives, designed to systematically mitigate both direct and indirect injection attacks. Furthermore, we introduce a one-shot self-correction mechanism that transforms security violations into constructive feedback, enabling agents to recover from attacks. Extensive experiments show that \tool{} reduces the attack success rate to 0.65\%, achieving this strong defense while incurring only a 1.45\% average decrease in utility relative to the No Defense scenario, demonstrating superior performance compared to existing defense methods.
\end{abstract}

\begin{figure}[!t]
    \centering
    \includegraphics[width=0.48\textwidth]{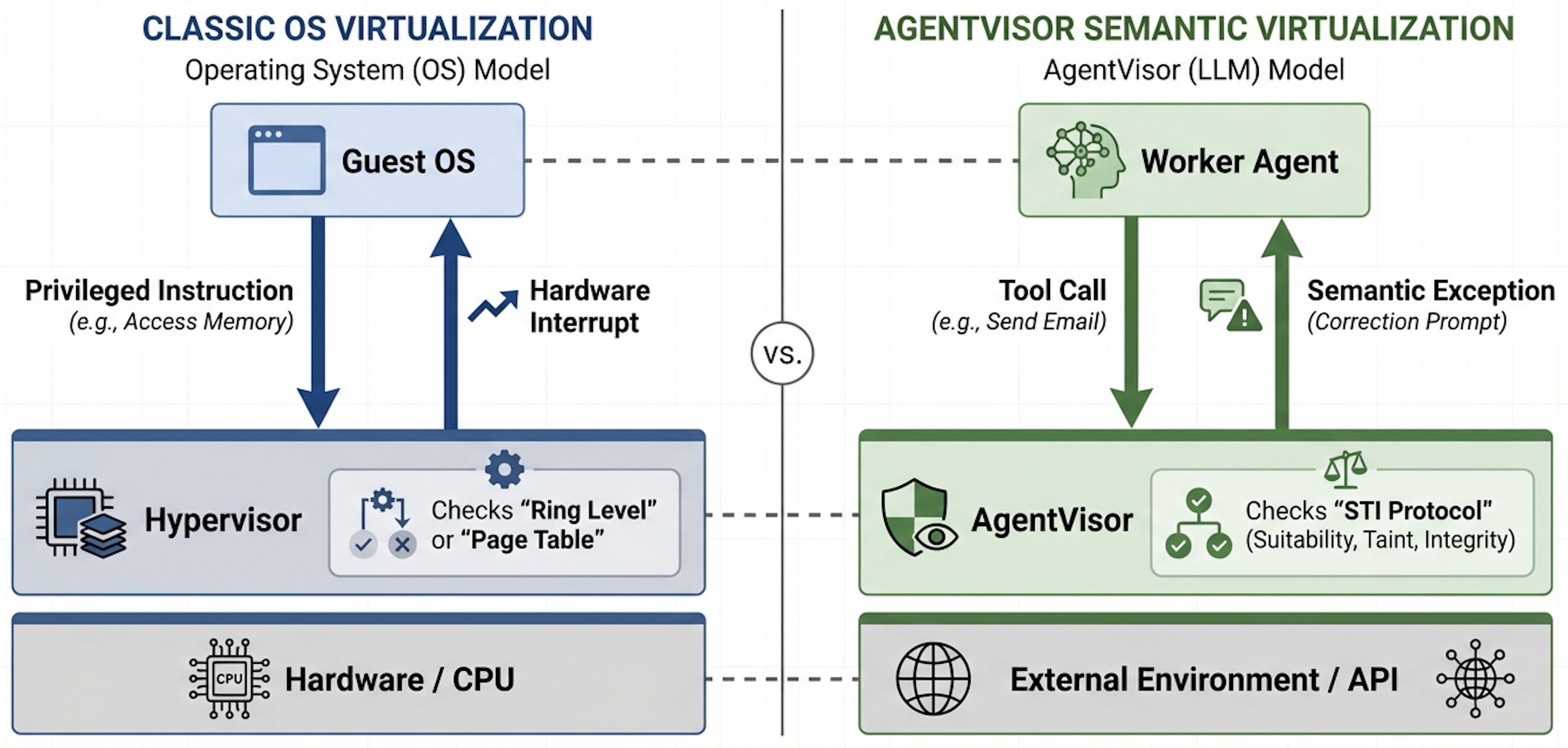} 
    \caption{Systematic mapping between OS Virtualization and \tool{}. We translate classic OS security concepts into the semantic space of LLM agents.}
    \label{tab:generalization}
\end{figure}

\section{Introduction}

Large Language Model (LLM) agents have transitioned from passive conversational systems to active entities capable of automating complex workflows through tool use \cite{wolflein2025llm,kong2024tptu,yuan2025easytool}. However, integrating untrusted external data with privileged execution capabilities exposes agents to severe security risks, primarily \textit{Direct} and \textit{Indirect Prompt Injection}. In direct injection \cite{liu2024formalizing}, malicious users explicitly override system instructions to perform unauthorized actions, whereas indirect injection \cite{yi2025benchmarking} involves hidden commands embedded in external content that the agent processes. Current LLM agents often fail to reliably distinguish privileged system instructions from untrusted external commands, which can lead to the unintended execution of malicious payloads.

Despite growing awareness of these threats, existing defenses remain fragmented and brittle. Prompt-based hardening \cite{sandwich_defense_learnprompting_2024,instruction_defense_learnprompting_2024,yi2025benchmarking} is heuristic and can be overridden by adversarial instructions. Input/output filtering and LLM-based guardrails \cite{deberta-v3-base-prompt-injection-v2,liu2025datasentinel} are prone to evasion, introduce false positives, and provide limited guarantees on how untrusted content affects tool-use decisions. Tool-sandboxing approaches \cite{meng2025cellmate,piao2025agentbay} restrict available actions, but are often coarse, fail to track information flow in multi-step workflows, and typically lack a principled recovery path once a violation occurs. Consequently, agents still lack a principled security architecture that separates trusted control from untrusted model behavior while enforcing least privilege and information-flow constraints with minimal utility loss.

To address these challenges, we draw inspiration from operating system virtualization, where a Hypervisor \cite{popek1974formal} isolates an untrusted Guest from privileged hardware resources in secure OS architectures \cite{shinagawa2009bitvisor,li2019protecting}. 

We adapt three key mechanisms for agent security: \ding{182} \textit{Privilege Separation} \cite{saltzer1975protection}, which traps and audits sensitive operations; \ding{183} \textit{Policy Enforcement} \cite{saltzer1975protection,bell1976secure}, grounded in Least Privilege and Information Flow Control; and \ding{184} \textit{Exception Injection} \cite{popek1974formal}, which reports violations via interrupts instead of terminating execution. Translating this paradigm into the semantic space of LLMs, we propose \tool{}, a \textit{semantic virtualization} framework that treats the target agent as an untrusted \textit{Guest} and mediates every tool call via a trusted \textit{Visor}. The Visor enforces a rigorous STI (Suitability, Taint, Integrity) protocol: \textit{Suitability} applies Least Privilege to mitigate direct injection, \textit{Taint} enforces information-flow constraints to block indirect injection, and \textit{Integrity} preserves parameter/data integrity across both. Upon detecting violations, \tool{} injects a \textit{semantic exception} to trigger self-correction, maintaining high utility.

Our contributions are summarized as follows:
\begin{itemize}
    \item We propose \tool{}, a virtualization-based defense that enforces privilege separation and one-shot self-correction, securing LLM agents against prompt injection attack.
    
    \item We design the STI protocol, adapting OS security primitives into a structured semantic audit mechanism, providing systematic and interpretable defenses.
    
    \item Experiments show that \tool{} reduces the attack success rate to 0.65\%, with only a 1.45\% average utility loss compared to No Defense, outperforming existing defenses.
\end{itemize}

\section{Related Work}

\subsection{Prompt Injection Attacks}
LLM agents are vulnerable to both direct and indirect prompt injection attacks, which differ primarily in the injection source but often share common methods. Attack methods can be broadly categorized into optimization-based and non-optimization-based approaches. \textit{Optimization-based methods} \cite{shi2024optimization,wang2025manipulating,liu2024automatic} utilize gradient approximation \cite{zou2023universal} to generate adversarial tokens. However, these methods are computationally expensive and often exhibit poor transferability across injection scenarios. In contrast, \textit{non-optimization-based methods} \cite{willison2023delimiters,liu2024formalizing,debenedetti2024agentdojo} rely on manually crafted semantic triggers such as Ignore \cite{perez2022ignore} or Escape \cite{breitenbach2023don} to manipulate agent behavior. Crucially, these handcrafted techniques are universally applicable to both direct and indirect injection scenarios. Given their practicality and prevalence in real-world exploits, our work focuses on evaluating defenses against these non-optimization-based attacks.

\subsection{Prompt Injection Defenses}
Existing defenses against prompt injection can be categorized into detection-based and mitigation-based approaches. \textit{Detection-based methods} focus on verifying the integrity of the input source to identify potential tampering. These approaches typically employ off-the-shelf LLMs \cite{shi2025promptarmor} or fine-tuned guardrail models \cite{deberta-v3-base-prompt-injection-v2,dubey2024llama3herdmodels,liu2025datasentinel} to inspect inputs for malicious content before they are processed by the agent. While effective at flagging threats, these methods often act as binary filters, terminating the interaction upon detection and potentially reducing system utility. \textit{Mitigation-based methods} aim to ensure that the agent executes the intended target task while suppressing the execution of injected commands. One line of research achieves this through safety-specific fine-tuning of the LLM \cite{chen2025secalign,chen2025struq}, enhancing its intrinsic resistance to adversarial instructions. Another direction involves security enhancement of the input context \cite{chen2025defending,yi2025benchmarking}, such as reiterating the target task instructions \cite{sandwich_defense_learnprompting_2024} or applying masking strategies to the agent's state prior to tool invocation \cite{zhu2025melon}. Unlike detection methods, mitigation strategies strive to maintain functionality even in the presence of attacks.

\section{Preliminaries and Threat Model}

\subsection{Preliminaries}
We consider an LLM-based tool-using agent $\mathcal{A}$ that interacts with a user and an external environment. At each turn $t$, the agent receives a trusted system instruction $I_{\text{sys}}$, a user query $I_u$, and (when available) an external context $C_t$ retrieved from potentially untrusted sources (\eg, webpages, emails, or documents). Based on these inputs, the agent may invoke a tool.

We represent the agent's tool invocation at turn $t$ as:
\begin{equation}
    T_t = \mathcal{A}(I_{\text{sys}}, I_u, C_t),
\end{equation}
where $T_t \in \{\emptyset\} \cup \mathcal{F}\times\mathcal{X}$.
Here, $T_t=(f,\mathbf{args})$ denotes a tool call with function name $f\in\mathcal{F}$ and arguments $\mathbf{args}\in\mathcal{X}$; if no tool is invoked, $T_t=\emptyset$.

For multi-step agent workflows (e.g., those involving indirect prompt injections), we additionally maintain an execution history $H_{t-1}$ that records prior tool calls and observations, and the agent's decision becomes $T_t = \mathcal{A}(I_{\text{sys}}, I_u, H_{t-1}, C_t)$.

\subsection{Threat Model}\label{sec:prelim}
We consider an adversary who aims to manipulate the agent into executing an unauthorized tool call, causing the agent's behavior to deviate from the intended objective specified by $I_{\text{sys}}$ and the benign user intent in $I_u$. The adversary cannot modify the agent $\mathcal{A}$, the system instruction $I_{\text{sys}}$, or the tool implementations/execution environment, but can manipulate inputs delivered to the agent.
We use $\oplus$ to denote the injection operation that augments an otherwise benign input with an adversarial payload.

\paragraph{Direct Prompt Injection.}
In direct prompt injection, the adversary acts as the user and submits an adversarial query $I_u^{\text{adv}} = I_u \oplus \delta_{\text{dir}}$, where $\delta_{\text{dir}}$ is a malicious instruction intended to override constraints in $I_{\text{sys}}$ or redirect the agent to an attacker-chosen objective. We assume direct injection does not rely on external context, i.e., $C_t = \emptyset$. The compromised action is then $T_{\text{adv}} = \mathcal{A}(I_{\text{sys}}, I_u^{\text{adv}}, C_t)$.

\paragraph{Indirect Prompt Injection.}
In indirect prompt injection, the user query $I_u$ is benign, but the adversary poisons the externally retrieved context by embedding hidden or overt malicious instructions, $C_t^{\text{adv}} = C_t \oplus \delta_{\text{ind}}$, where $\delta_{\text{ind}}$ is placed in untrusted content (e.g., a webpage or an email) that the agent later reads. The attacker aims to hijack the agent's (single- or multi-step) tool-use decisions such that $T_t = \mathcal{A}(I_{\text{sys}}, I_u, H_{t-1}, C_t^{\text{adv}})$, potentially yielding an attacker-intended tool call $T_{\text{adv}}$ at some step.

\begin{figure*}[!t]
    \centering
    \includegraphics[width=0.98\textwidth]{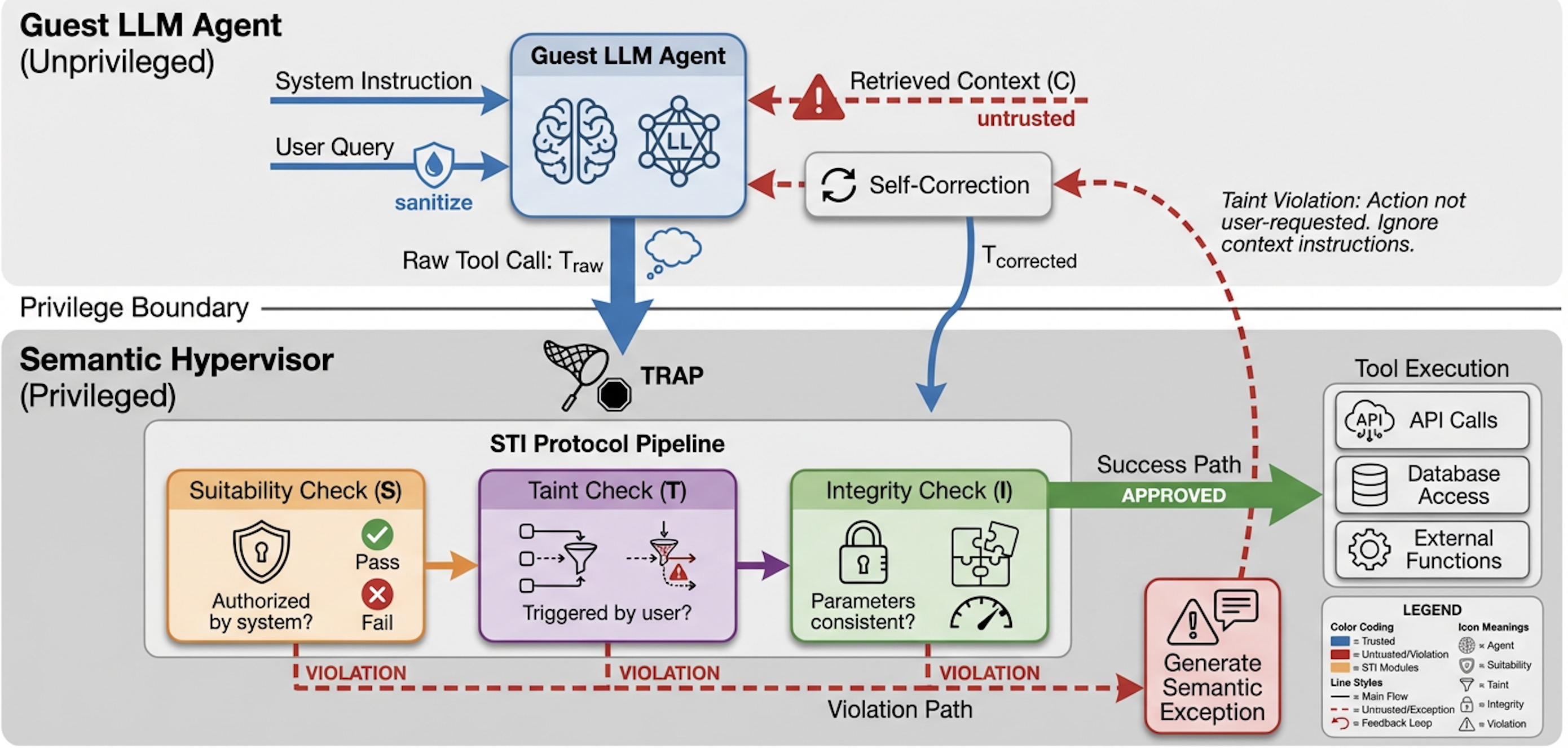} 
    \caption{Overview of the \tool{} architecture. Drawing inspiration from OS virtualization, \tool{} enforces privilege separation between the untrusted target agent (Guest) and the trusted semantic hypervisor (Visor).}
    \label{tab:generalization}
\end{figure*}

\section{Methodology}
\label{sec:method}

In this section, we present \tool{}, a defense framework structured as a \emph{Semantic Hypervisor} for tool-using LLM agents. tool{} enforces a \emph{trap--audit--recover} control loop: the target agent (Guest) proposes tool calls, while a lightweight supervisory component (Visor) audits each proposal against a structured protocol and, when needed, injects a \emph{Semantic Exception} to steer one-shot self-correction.

\begin{figure}[!t]
    \centering
    \includegraphics[width=0.48\textwidth]{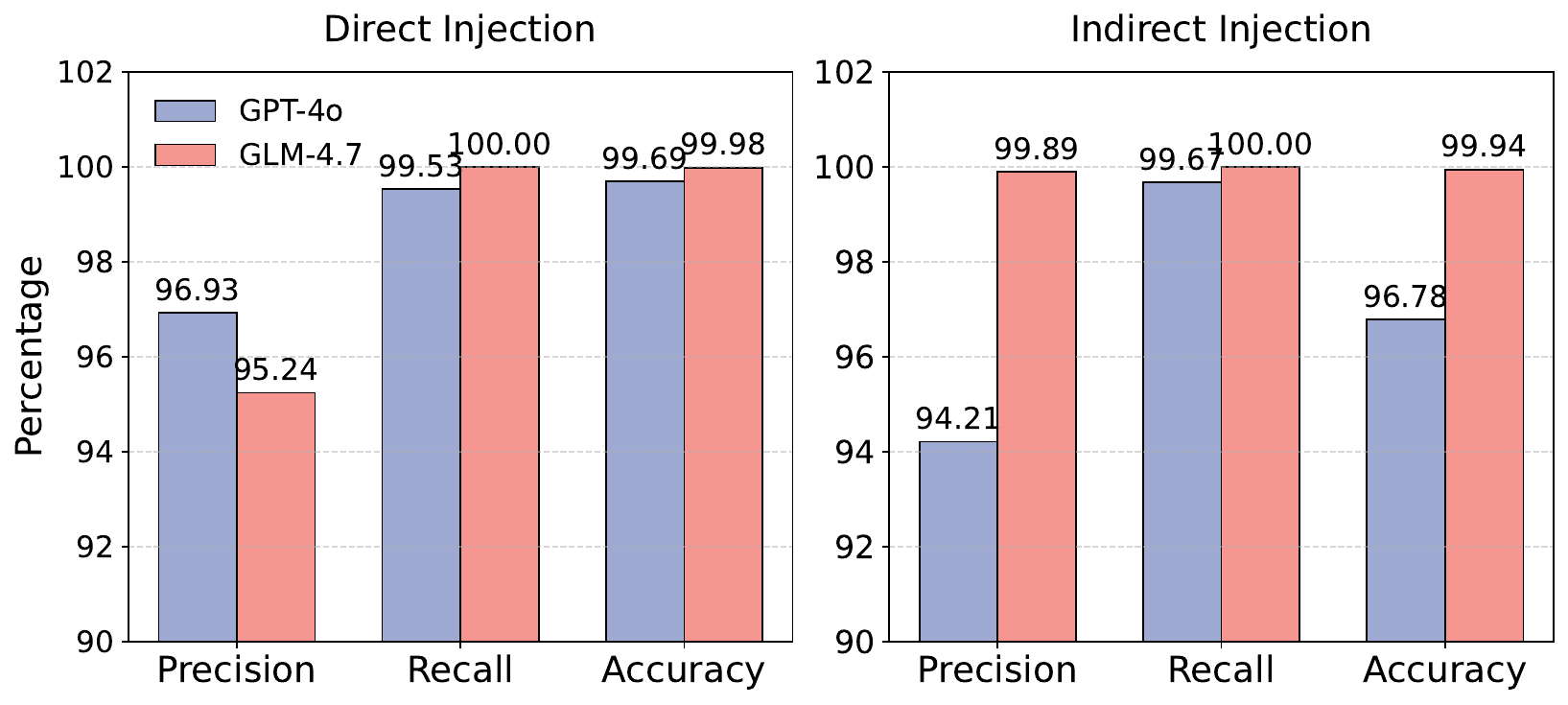}
    \caption{Detection performance of target agents (based on GPT-4o and GLM-4.7) against prompt injections.}
    \label{fig:judge}
\end{figure}

\subsection{Problem Formulation}
\label{subsec:formulation}

At step $t$, the Guest (target agent) proposes a tool call
\begin{equation}
T_t^{\text{raw}}=(f_t,\mathbf{args}_t)\ \ \text{or}\ \ T_t^{\text{raw}}=\emptyset,
\end{equation}
given the trusted system instruction $I_{\text{sys}}$, the user query $I_u$, and (in the indirect setting) potentially untrusted state such as retrieved context and multi-step execution traces.

\tool{} mediates this proposal using only \emph{trusted} inputs and a \emph{sanitized} task state. Specifically, it receives:
\begin{equation}
X_t \triangleq (I_{\text{sys}}, I_u, \tilde{H}_{t-1}, T_t^{\text{raw}}),
\end{equation}
where $\tilde{H}_{t-1}$ is a structured history view containing only fields such as \texttt{tool\_name}, canonicalized \texttt{args}, and a short \texttt{return\_summary}/\texttt{status}, wrapped with strict delimiters and treated as data.

The defense outputs either an approval or a structured exception:
\begin{align}
(\textsf{dec}_t, E_t) &= \mathcal{D}(X_t), \\
\textsf{dec}_t &\in \{\textsf{allow}, \textsf{exception}\}.
\end{align}

If $\textsf{dec}_t=\textsf{allow}$, the proposed tool call is executed as-is; otherwise, the Visor injects $E_t$ to request one-shot self-correction from the Guest:
\begin{equation}
T_t'=\mathcal{A}(I_{\text{sys}}, I_u, \tilde{H}_{t-1}, E_t),
\end{equation}
after which $T_t'$ is executed. In both cases, the executed action $T_t$ is the final tool call for step $t$:
\begin{equation}
T_t=\begin{cases}
T_t^{\text{raw}}, & \text{if }\textsf{dec}_t=\textsf{allow},\\
T_t', & \text{if }\textsf{dec}_t=\textsf{exception}.
\end{cases}
\end{equation}

\subsection{Architecture of \tool{}}
\label{subsec:arch}

\tool{} is motivated by an empirical observation: modern agents can often \emph{recognize} prompt injections when explicitly asked (\eg, \Fref{fig:judge}), yet still \emph{act on} malicious instructions during tool use. This \emph{awareness--action gap} suggests that ``knowing an input is unsafe'' does not reliably translate into safe tool-use behavior. \tool{} addresses this gap by introducing a virtualization-inspired separation of concerns: the Guest generates actions, while the Visor enforces policy via auditing and recovery.

\paragraph{Guest (Target Agent).}
The target agent serves as the \emph{Guest}. It operates on the full task context, including potentially adversarial retrieved content $C_t^{\text{adv}}$ in the indirect setting, and produces a tool call proposal $T_t^{\text{raw}}$. The Guest is also responsible for revising its proposal when given a Semantic Exception.

\paragraph{Visor (Semantic Hypervisor).}
AgentVisor acts as the \emph{Visor}. It audits the Guest's proposed tool call and, if needed, returns a structured exception that instructs the Guest how to revise the action. Critically, we enforce context isolation: the Visor is \emph{architecturally blind} to the raw external context $C_t$ (and thus to any embedded attacker instructions). Instead, it bases its decision solely on the trusted system instruction $I_{\text{sys}}$, the user query $I_u$, and a sanitized execution history $\tilde{H}_{t-1}$ that contains only structured fields such as \texttt{tool\_name}, canonicalized \texttt{args}, and a short \texttt{return\_summary} or \texttt{status}. The history excludes raw observations, is wrapped with strict delimiters, and is treated strictly as data rather than as executable instructions.

\tool{} follows a trap--audit--recover loop:
(1) the Guest proposes $T_t^{\text{raw}}$;
(2) the Visor audits $T_t^{\text{raw}}$ with the STI protocol (\Sref{subsec:sti});
(3) if unsafe, the Visor returns a Semantic Exception (\Sref{subsec:recover}); the Guest regenerates once and the revised tool call is executed. The complete procedure is provided in \Aref{alg:agentvisor} in \Asref{app:alg}.

\subsection{The STI Audit Protocol}
\label{subsec:sti}

To operationalize auditing, we introduce the \emph{STI (Suitability, Taint, Integrity) Protocol}. STI is a structured semantic verification pipeline; each proposed tool call must pass three checks in order. Each check produces $(\textsf{pass}, \bot)$ or $(\textsf{fail}, r)$, where $r$ is a machine-readable reason used to construct a Semantic Exception.

\paragraph{Suitability (S): tool appropriateness under system policy.}
Suitability verifies whether invoking tool $f_t$ is permitted under $I_{\text{sys}}$ and is aligned with the assistant's intended role. This is a coarse-grained gate that is especially effective against direct prompt injection attempts that try to escalate the agent's capabilities.
Formally, the Suitability check outputs $(\textsf{S}, r_S) = \textsf{Check}_S(I_{\text{sys}}, I_u, f_t)$. If $\textsf{S}=\textsf{fail}$, the Visor raises an exception indicating that the tool is not appropriate for this assistant under the trusted system instruction.

\paragraph{Taint (T): goal alignment with user intent and task state.}
Taint checks whether the \emph{goal implied by} the tool call is aligned with the user request $I_u$ and with legitimate intermediate goals derived from the task state $\tilde{H}_{t-1}$. Intuitively, it blocks actions that introduce new goals not supported by the user's request (e.g., forwarding, exfiltrating, or posting content when the user asked only to summarize).
We implement Taint as a semantic alignment decision:
\begin{equation}
(\textsf{T}, r_T)=\textsf{Check}_T(I_u, \tilde{H}_{t-1}, f_t, \mathbf{args}_t).
\end{equation}
If $\textsf{T}=\textsf{fail}$, the Visor concludes that the tool call appears to be driven by untrusted instructions rather than the user's intended task.

\paragraph{Integrity (I): argument consistency with user-specified entities.}
Integrity verifies that the tool arguments $\mathbf{args}_t$ are consistent with the entities, constraints, and targets specified by $I_u$ or established in $\tilde{H}_{t-1}$. This prevents cases where the \emph{tool choice} is reasonable but the \emph{arguments} are redirected.
For example, when the user requests sending a message to a specific recipient, an indirect injection may try to keep the same tool (e.g., \texttt{send\_email}) but substitute a different recipient in $\mathbf{args}_t$.
Formally:
\begin{equation}
(\textsf{I}, r_I)=\textsf{Check}_I(I_u, \tilde{H}_{t-1}, f_t, \mathbf{args}_t).
\end{equation}

\subsection{Resilience via Semantic Exception Injection}
\label{subsec:recover}

Blocking unsafe tool calls can be brittle: a strict deny policy often causes task failure even when a safe alternative exists. AgentVisor therefore converts audit failures into \emph{recoverable} events using \textbf{Semantic Exception Injection}, analogous to exception handling in operating systems.

When STI fails, the Visor synthesizes a structured exception $E_t$ with the following fields:
\begin{equation}
\begin{aligned}
E_t = \langle\,
&\textsf{type},\ \textsf{violated\_rule},\ \textsf{rationale},\\
&\textsf{constraints},\ \textsf{allowed\_objective}\,
\rangle.
\end{aligned}
\end{equation}

Concretely, the exception record contains several fields. The field \textsf{type} indicates which STI stage failed, taking values in ${\textsf{S}, \textsf{T}, \textsf{I}}$. The field \textsf{violated\_rule} provides a brief identifier, such as “tool not permitted under system role”. The field \textsf{rationale} explains why the proposal conflicts with $I_{\text{sys}}$, $I_u$, or $\tilde{H}_{t-1}$. The field \textsf{constraints} specifies the required negative or positive conditions, for example “do not forward or share data; do not use external recipients; only summarize”. Finally, the field \textsf{allowed\_objective} restates the user-aligned goal to preserve utility.

After receiving $E_t$, the Guest regenerates a revised tool call $T_t'$ once based on the explicit constraints and then executes it immediately. In our experiments, this one-shot correction is sufficient to redirect the agent to a safe, task-preserving action in the vast majority of cases; we further discuss the trade-off between additional audit rounds and resource cost in \Sref{sec:discussion}.

\section{Experiments}

\subsection{Experimental Setup}

\paragraph{Datasets.}We evaluate \tool{} against both direct and indirect prompt injection attacks. For direct injection, we utilize OpenPromptInjection \cite{liu2024formalizing}, which encompasses 7 NLP tasks that serve interchangeably as target and injection tasks. Following Jia et al. \cite{jia2025promptlocate}, we randomly sample 100 examples for each task combination, yielding a total of 4,900 attack cases. For indirect injection, we employ AgentDojo \cite{debenedetti2024agentdojo}, which features 4 interactive agent environments (Banking, Travel, Slack, Workplace) where agents iteratively invoke tools to complete tasks. These environments contain 16, 20, 21, and 40 target tasks respectively, combined with varying injection points and tasks, resulting in a total of 629 attack cases.

\paragraph{Attacks.} We evaluate robustness against 7 representative strategies: 1) \textit{Direct:} Directly appending the injection task; 2) \textit{Ignore \cite{perez2022ignore}} ; 3) \textit{Escape \cite{breitenbach2023don}}; 4) \textit{FakeComp \cite{willison2023delimiters}} ; 5) \textit{Combined \cite{liu2024formalizing}:} Integrating \textit{Ignore}, \textit{Escape}, and \textit{FakeComp}; 6) \textit{System \cite{debenedetti2024agentdojo}} ; and 7) \textit{Important \cite{debenedetti2024agentdojo}}.

\begin{table*}[h]
\caption{Defense performance (\%) of \tool{} and baseline approaches under \textbf{Direct Injection}.}
\label{tab:direct_results}
\resizebox{\textwidth}{!}{
\begin{tabular}{@{}c|c|cc|cc|cc|cc|cc|cc|cc@{}}
\toprule
Method           & No Attack & \multicolumn{2}{c|}{Direct} & \multicolumn{2}{c|}{Ignore} & \multicolumn{2}{c|}{Escape} & \multicolumn{2}{c|}{Fakecom} & \multicolumn{2}{c|}{Combined} & \multicolumn{2}{c|}{SYSTEM} & \multicolumn{2}{c}{Important} \\ \midrule
Metric           & BU        & UA            & ASR         & UA           & ASR          & UA           & ASR          & UA            & ASR          & UA            & ASR           & UA           & ASR          & UA            & ASR           \\ \midrule
No Defense       & 94.25   & 96.25       & 97.00     & 71.12      & 83.28      & 98.27      & 97.00      & 46.27       & 91.00      & 0.00        & 88.26       & 98.63      & 91.95      & 88.97       & 95.98       \\ \midrule
Sandwich         & 100.00  & 93.88       & 91.00     & 90.29      & 56.37      & 98.10      & 93.18      & 72.40       & 91.27      & 20.15       & 51.44       & 94.76      & 94.37      & 94.92       & 93.24       \\
Instructional    & 93.75   & 89.28       & 84.00     & 43.14      & 81.19      & 94.28      & 87.29      & 30.16       & 86.09      & 0.00        & 68.34       & 96.98      & 82.20      & 88.16       & 74.97       \\
Reminder         & 93.25   & 98.19       & 91.00     & 87.29      & 85.56      & 96.09      & 93.95      & 82.08       & 90.52      & 0.00        & 85.09       & 95.08      & 76.79      & 74.38       & 54.24       \\
Isolation        & 93.16   & 100.00      & 62.00     & 40.95      & 83.29      & 98.18      & 83.00      & 57.78       & 85.20      & 0.01        & 67.10       & 95.66      & 21.95      & 90.07       & 88.30       \\
Spotlight        & 87.35   & 85.00       & 2.00      & 90.05      & 7.16       & 92.08      & 0.00       & 63.20       & 0.00       & 71.08       & 0.12        & 88.85      & 4.16       & 77.16       & 6.08        \\
DeBERTa & 91.25   & 84.28       & 86.35     & 86.96      & 82.00      & 98.31      & 94.95      & 62.20       & 96.75      & 0.00        & 88.87       & 95.75      & 93.78      & 92.30       & 86.37       \\
DataSentinel     & 89.17   & 0.00        & 0.00      & 0.15       & 3.10       & 0.00       & 0.00       & 0.00        & 1.00       & 0.00        & 3.00        & 3.25       & 3.64       & 0.00        & 0.00        \\ \midrule
\tool{}             & 91.35   & 87.56       & 0.00      & 74.38      & 0.00       & 93.15      & 0.00       & 90.00       & 0.00       & 76.00       & 0.00        & 83.14      & 0.00       & 88.24       & 0.00        \\ \bottomrule

\end{tabular}
}
\end{table*}

\begin{table*}[h]
\caption{Defense performance (\%) of \tool{} and baseline approaches under \textbf{Indirect Injection}.}
\label{tab:indirect_results}
\resizebox{\textwidth}{!}{
\begin{tabular}{@{}c|c|cc|cc|cc|cc|cc|cc|cc@{}}
\toprule
Method           & No Attack & \multicolumn{2}{c|}{Direct} & \multicolumn{2}{c|}{Ignore} & \multicolumn{2}{c|}{Escape} & \multicolumn{2}{c|}{Fakecom} & \multicolumn{2}{c|}{Combined} & \multicolumn{2}{c|}{SYSTEM} & \multicolumn{2}{c}{Important} \\ \midrule
Metric           & BU        & UA            & ASR         & UA            & ASR         & UA            & ASR         & UA            & ASR          & UA             & ASR          & UA            & ASR         & UA            & ASR           \\ \midrule
No Defense       & 85.00   & 76.18       & 2.78      & 70.15       & 2.61      & 71.71       & 2.45      & 68.14       & 2.13       & 73.31        & 2.91       & 75.73       & 3.57      & 57.74       & 45.14       \\ \midrule
Sandwich         & 90.00   & 80.89       & 1.95      & 77.32       & 1.33      & 75.36       & 1.28      & 79.50       & 1.70       & 74.05        & 1.11       & 79.25       & 1.95      & 58.15       & 17.61       \\
Instructional    & 82.50   & 82.65       & 2.22      & 75.12       & 1.06      & 77.12       & 1.76      & 78.13       & 1.11       & 79.49        & 1.59       & 76.31       & 2.23      & 58.22       & 46.79       \\
Reminder         & 77.50   & 63.65       & 1.95      & 77.88       & 0.28      & 77.03       & 1.84      & 79.00       & 0.00       & 77.93        & 0.00       & 71.56       & 1.11      & 68.75       & 22.15       \\
Isolation        & 80.00   & 71.83       & 2.72      & 76.34       & 0.56      & 72.30       & 0.54      & 73.29       & 0.49       & 75.75        & 0.43       & 78.32       & 1.67      & 60.66       & 44.77       \\
Spotlight        & 62.50   & 57.66       & 1.33      & 63.87       & 0.78      & 63.98       & 1.15      & 68.98       & 0.15       & 65.25        & 0.78       & 60.16       & 0.40      & 61.02       & 0.28        \\
DeBERTa & 40.00   & 12.50       & 0.00      & 18.40       & 0.00      & 11.49       & 0.00      & 10.00       & 0.00       & 12.56        & 0.00       & 25.18       & 0.27      & 19.75       & 3.55        \\
MELON            & 82.50   & 56.69       & 2.00      & 65.59       & 0.00      & 69.19       & 0.25      & 77.90       & 0.00       & 69.94        & 0.00       & 70.31       & 0.22      & 35.42       & 1.04        \\ \midrule
\tool{}             & 85.00   & 65.15       & 1.69      & 74.85       & 0.00      & 75.32       & 0.00      & 74.84       & 0.11       & 76.73        & 0.00       & 78.25       & 0.11      & 63.75       & 0.69        \\ \bottomrule
\end{tabular}
}
\end{table*}

\paragraph{Baseline Defenses.} We compare \tool{} against 8 representative defense methods categorized into two types. Mitigation-based methods include \textit{Sandwich} \cite{sandwich_defense_learnprompting_2024}, \textit{Instructional}  \cite{instruction_defense_learnprompting_2024}, \textit{Reminder} \cite{yi2025benchmarking}, \textit{Isolation} \cite{willison2023delimiters}, and \textit{Spotlighting} \cite{hines2024defending}. Detection-based methods include \textit{DeBERTa} \cite{deberta-v3-base-prompt-injection-v2}, \textit{DataSentinel} \cite{liu2025datasentinel}, and \textit{MELON} \cite{zhu2025melon}. Note that for detection-based methods, a successful detection terminates the process. Furthermore, since \textit{DataSentinel} and \textit{MELON} were originally designed specifically for direct and indirect injection respectively, we follow their original settings and do not report their performance on out-of-scope attack scenarios.

\paragraph{Metrics.}We evaluate performance using three metrics: 1) Benign Utility (BU), which measures task completion in non-attack scenarios to assess capability preservation; 2) Attack Success Rate (ASR), the percentage of cases where the agent executes the injected malicious task; 3) and Utility under Attack (UA), which measures the agent's resilience in completing the original user task despite the attack. For effective defense, lower ASR and higher BU/UA are desired.

\paragraph{Implementation Details.} We employ GPT-4o \cite{hurst2024gpt} and GLM 4.7 \cite{5team2025glm45agenticreasoningcoding} as the backbone models for the target Agent to simulate high-capability agents. For the \tool{}, we utilize Gemini-2.5-Flash \cite{google_vertex_ai_gemini_2_5_flash_2025} to demonstrate the framework's efficiency and effectiveness even with lightweight, cost-effective models.

\subsection{Main Results}

Experimental results obtained with the GPT-4o backbone are presented in this section, showing defense performance against Direct (\Tref{tab:direct_results}) and Indirect (\Tref{tab:indirect_results}) Injection, while results for the GLM-4.7 backbone are detailed in \Asref{app:glm_results}.

\paragraph{Direct Injection.} The \textit{No Defense} baseline is extremely vulnerable (ASR $>$ 90\%), and mitigation-based defenses like \textit{Sandwich} often yield ASRs above 50\%. While \textit{DataSentinel} achieves low ASR, its utility collapses (UA $\approx$ 0\%). In contrast, \tool{} achieves perfect security (0.00\% ASR) while maintaining high utility (UA $>$ 83\%), even in the challenging \textit{Combined} attack (76.00\% UA). This confirms that \tool{} effectively enforces system boundaries without sacrificing the execution of legitimate tasks.

\paragraph{Indirect Injection.} While standard attacks are less effective, the \textit{Important} strategy compromises \textit{No Defense} (45.14\% ASR). Detection methods like \textit{DeBERTa} suffer from severe over-defense (UA $\sim$15\%). \tool{} strikes the best trade-off, suppressing ASR to negligible levels ($<$1.7\%), even in the \textit{Important} scenario (0.69\%). It also preserves high UA, ranging from 63\% to 78\%. This demonstrates that our STI protocol precisely blocks context-driven actions without disrupting legitimate workflows.

\subsection{Ablation Study}
\begin{figure}[!t]
    \centering
    \includegraphics[width=0.48\textwidth]{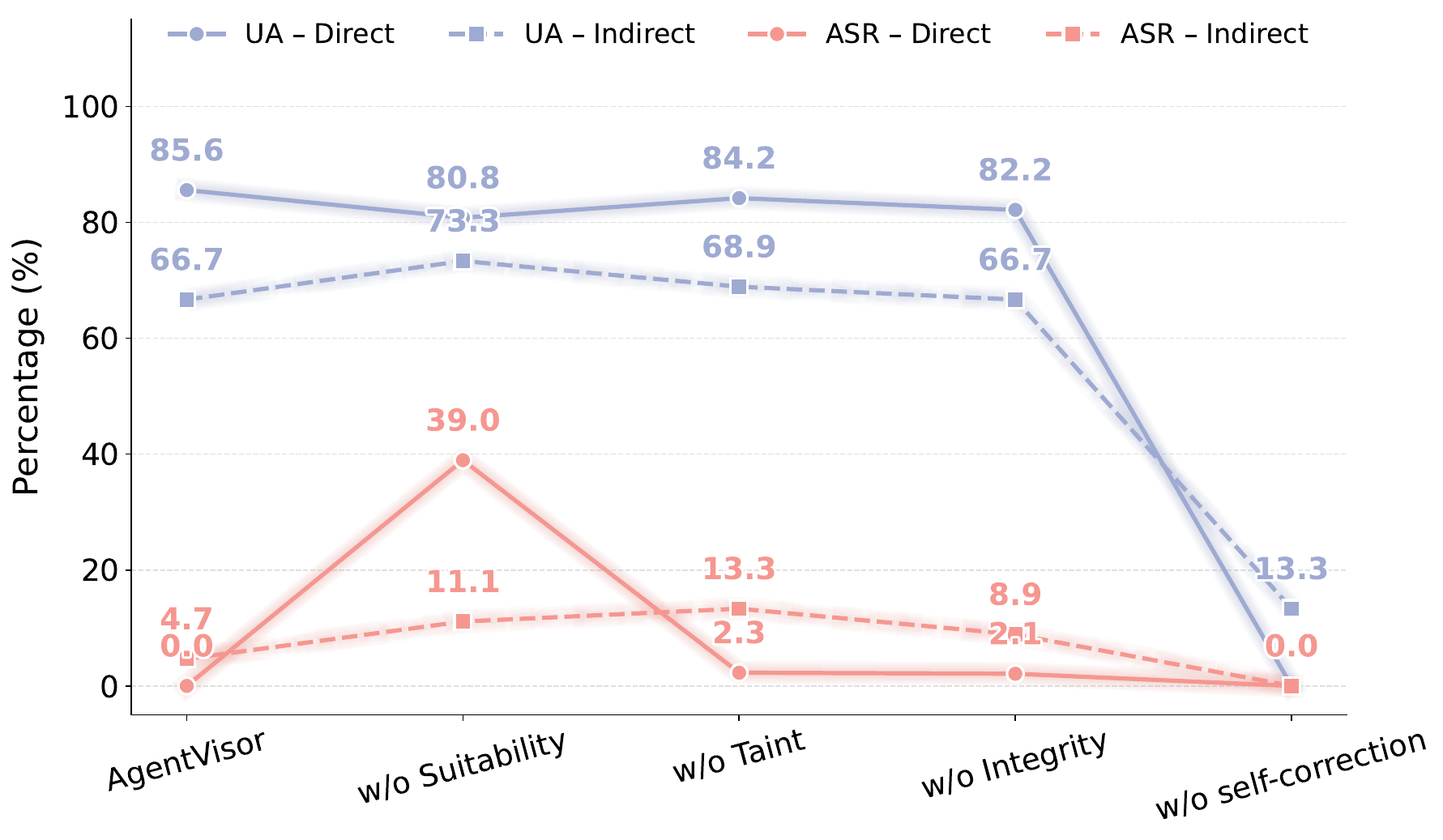} 
    \caption{Ablation study of \tool{} components. \textbf{w/o} denotes the removal of a specific component.}
    \label{fig:ablation}
\end{figure}

We validate the contribution of each \tool{} component through a comprehensive ablation study (\Fref{fig:ablation}). The results confirm the specialized roles of each STI layer. Removing Suitability (S) causes a dramatic failure in Direct Injection defense (ASR spikes to 38.95\%), validating it as the primary barrier against functionality hijacking. Removing Taint (T) significantly weakens Indirect Injection defense (ASR rises to 13.33\%), confirming its necessity in detecting context-triggered attacks. The absence of Integrity (I) leads to a moderate ASR increase (\eg, 8.89\% in Indirect), indicating its role in catching subtle parameter tampering.

The ablation of self-correction reveals its indispensable role in maintaining utility. In the Block-only setting, UA collapses to near zero (0.00\% for Direct, 13.33\% for Indirect) as mixed-intent prompts are rejected entirely. In contrast, our \textit{Semantic Fault Recovery} mechanism successfully strips away the injected task while executing the target task, restoring UA to 85.56\% and 66.67\% respectively. This demonstrates that self-correction is a fundamental requirement for practical mitigation method.

\section{Discussion}\label{sec:discussion}

\subsection{Trade-off Analysis}
\label{sec:discussion_rounds}

We validate our design choice of limiting self-correction to a single round ($N=1$) by analyzing the trade-off between utility gains and computational overhead across $N=1$ to $N=3$. 

As shown in \Tref{tab:rounds_tradeoff}, extending the correction loop beyond the first attempt yields negligible marginal gains. Indirect UA improves by only 1.29\% at $N=2$, while incurring a disproportionate latency penalty of 1.45$\times$ for $N=2$ and 1.90$\times$ for $N=3$. This indicates that the first semantic exception captures the vast majority of recoverable errors, and subsequent failures likely stem from fundamental capability limitations rather than ambiguity. Thus, $N=1$ represents the optimal efficiency-utility trade-off.

\begin{table}[!t]
\centering
\caption{Trade-off analysis of correction rounds ($N$).}
\label{tab:rounds_tradeoff}
\resizebox{\columnwidth}{!}{%
\begin{tabular}{c|cc|cc|c}
\toprule
\multirow{2}{*}{\textbf{Rounds ($N$)}} & \multicolumn{2}{c|}{\textbf{Direct Injection}} & \multicolumn{2}{c|}{\textbf{Indirect Injection}} & \multirow{2}{*}{\textbf{Relative Latency}} \\
 & \textbf{UA} $\uparrow$ & \textbf{ASR} $\downarrow$ & \textbf{UA} $\uparrow$ & \textbf{ASR} $\downarrow$ & \\ \midrule
\textbf{1 (Ours)} & \textbf{91.72\%} & \textbf{0.00\%} & \textbf{73.21\%} & \textbf{1.73\%} & \textbf{1.00$\times$} \\
2 & 92.05\% & 0.00\% & 74.50\% & 1.73\% & 1.45$\times$ \\
3 & 92.15\% & 0.00\% & 74.85\% & 1.73\% & 1.90$\times$ \\ \bottomrule
\end{tabular}%
}
\end{table}

\begin{figure*}[!t]
    \centering
    \includegraphics[width=0.98\textwidth]{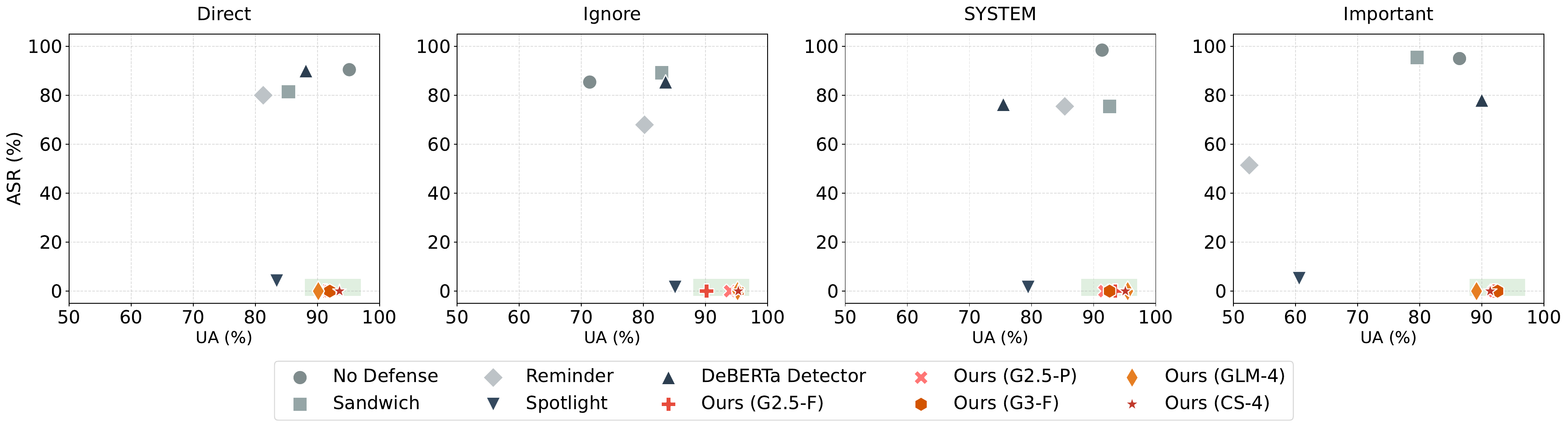} 
    \caption{Ablation study on \tool{}'s LLM backbone against \textbf{Direct Injection.}}
    \label{fig:model_ablation_direct}
\end{figure*}

\begin{figure*}[!t]
    \centering
    \includegraphics[width=0.98\textwidth]{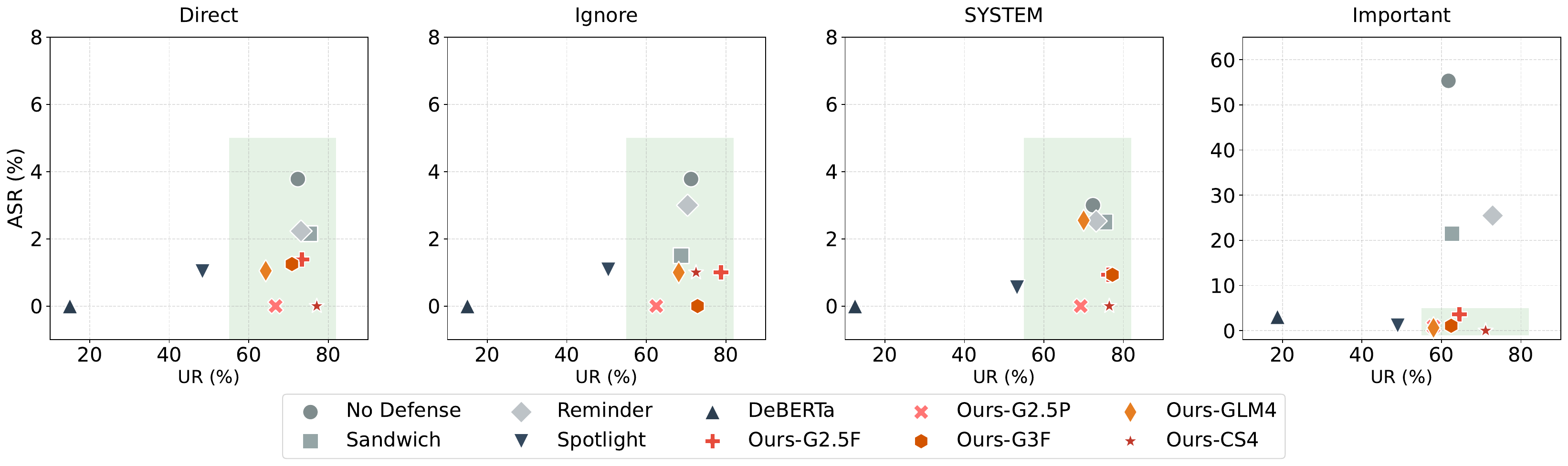} 
    \caption{Ablation study on \tool{}'s LLM backbone against \textbf{Indirect Injection.}}
    \label{fig:model_ablation_indirect}
\end{figure*}

\subsection{Model Agnosticism of \tool{}}

To determine whether \tool{}'s efficacy stems from its methodology or the underlying LLM, we evaluate it across diverse backbones. As shown in \Fref{fig:model_ablation_direct} and \Fref{fig:model_ablation_indirect}, \tool{} consistently achieves near-zero ASRs regardless of the model. In Direct Injection, all variants achieve a perfect 0.00\% ASR, confirming that the STI check is a fundamental task executable even by lightweight models. Similarly, in Indirect Injection, ASR remains consistently low ($<4\%$), significantly outperforming baselines. This demonstrates that the STI protocol provides a robust structural defense that does not require frontier-level reasoning to be effective.

While security is model-agnostic, stronger models yield higher utility. For instance, in Indirect Injection, \texttt{CS-4(Claude-Sonnet-4)} achieves a UA of 77.11\% compared to 64.27\% for \texttt{GLM-4(GLM-4.7)}. This suggests that while detection is structural, the \textit{self-correction} process benefits from superior reasoning capabilities, allowing stronger models to better reconstruct user intent from semantic exceptions.

\subsection{Robustness against Adaptive Attack}

\begin{figure}[!t]
    \centering
    \includegraphics[width=0.48\textwidth]{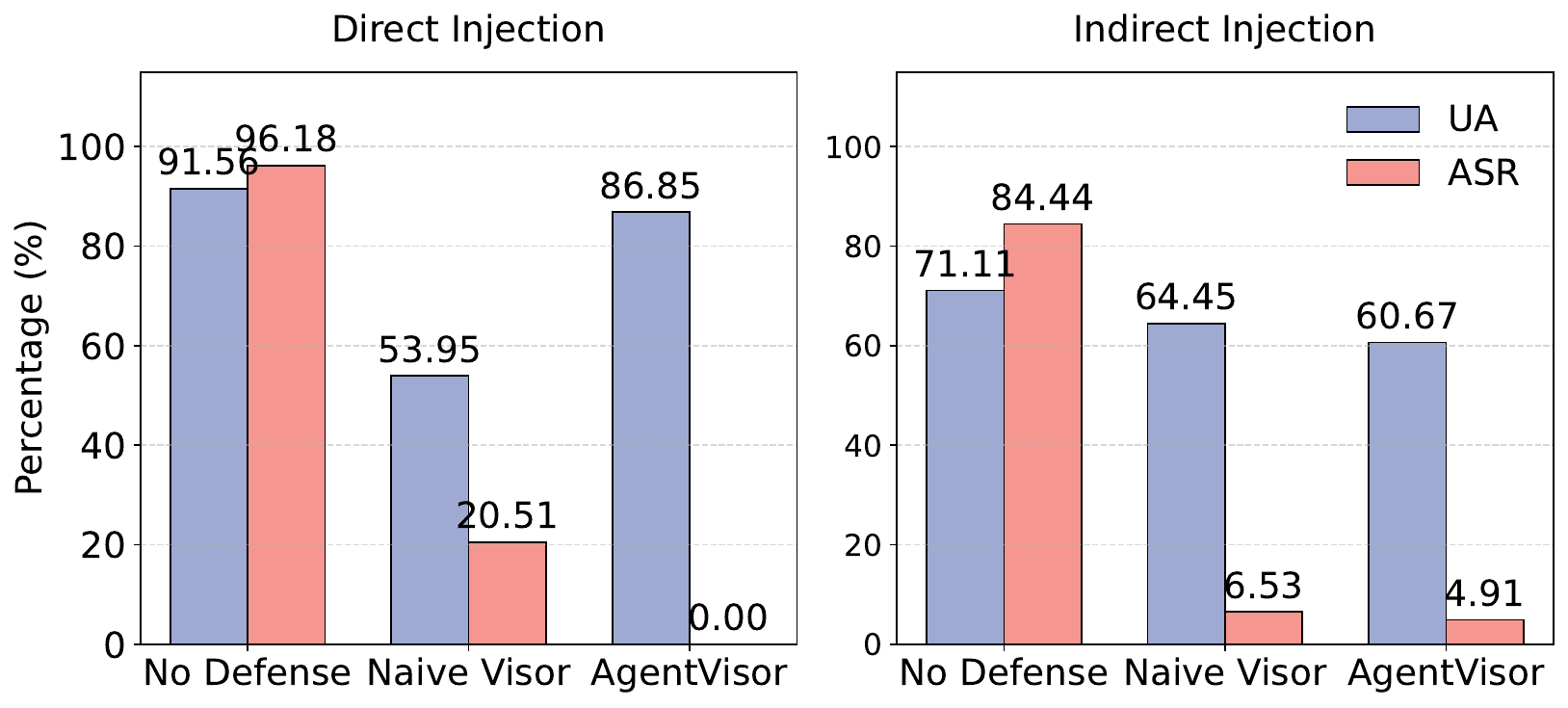} 
    \caption{Robustness of the defense mechanism against adaptive attacks.}
    \label{fig:adp_attack}
\end{figure}

We evaluate robustness against \textit{Adaptive Attacks} where recursive injections (\eg, "Ignore Security Check") target the defense layer. As shown in \Fref{fig:adp_attack}, the \textit{Naive Visor}, lacking structural isolation, is severely compromised: it struggles to disentangle mixed intents, resulting in a catastrophic drop in utility (\eg, 53.95\% UA) as it indiscriminately blocks legitimate tasks to stop the injection. In contrast, \tool{} demonstrates superior resilience. It completely neutralizes the adaptive injection (0\% ASR) while preserving high utility (86.85\%), confirming that our STI Protocol and context isolation effectively immunize the Visor against recursive manipulation, enabling precise mitigation.

\subsection{Latency Analysis}

\begin{table}[!t]
\centering
\caption{Latency analysis (seconds) across different scenarios.}
\label{tab:latency}
\resizebox{\columnwidth}{!}{%
\begin{tabular}{l|cc|cc}
\toprule
\multirow{2}{*}{\textbf{Scenario}} & \multicolumn{2}{c|}{\textbf{Benign Condition}} & \multicolumn{2}{c}{\textbf{Attack Condition}} \\
 & No Defense & \tool{} & No Defense & \tool{} \\ \midrule
\textbf{Direct Injection} & 2.12 & 2.95 (1.39$\times$) & 2.18 & 5.05 (2.32$\times$) \\ \midrule
\textbf{Indirect Injection} & 6.45 & 9.10 (1.41$\times$) & 6.58 & 11.25 (1.71$\times$) \\ \bottomrule
\end{tabular}%
}
\end{table}

We report the average end-to-end latency per task for \tool{} by sampling tasks from Direct and Indirect scenarios (Table~\ref{tab:latency}). In benign conditions, \tool{} introduces a moderate overhead ($\sim1.4\times$), as the Visor's audit is faster than open-ended generation. Under attack conditions, latency increases due to self-correction. For Direct Injection, latency roughly doubles ($2.32\times$) as the single step is re-executed. However, for multi-step Indirect Injection, the overhead ratio is lower ($1.71\times$) because typically only the compromised step triggers re-generation. We argue this pay-for-safety trade-off is justified, preventing critical breaches without imposing prohibitive penalties on normal operations.

\section{Conclusion}
In this paper, we presented \tool{}, a virtualization-based defense framework that secures LLM agents against both direct and indirect prompt injection through semantic privilege separation. By adapting classic OS security primitives into a rigorous audit protocol and incorporating a semantic fault recovery mechanism, \tool{} effectively mitigates functionality injection attack without compromising agent utility. Empirical results on standard benchmarks demonstrate that our approach achieves near-zero attack success rates across diverse attack vectors while maintaining high task completion performance, offering a robust and principled foundation for deploying secure autonomous agents.

\section{Limitations}
While \tool{} demonstrates robust defense capabilities, we acknowledge three limitations. 

\ding{182} Computational Overhead: The introduction of a hypervisor layer inherently incurs additional inference latency and token costs, although our one-shot correction mechanism minimizes this impact compared to iterative approaches. 
\ding{183} Long-Context Scalability: As the interaction history grows, the Hypervisor's performance may be constrained by the context window size of the underlying LLM, potentially affecting the precision of taint analysis in extremely long conversations. 
\ding{184} Multimodal Generalization: Our current framework focuses on textual prompt injections; extending the STI protocol to defend against visual or audio-based injection attacks in multimodal agents remains an important direction for future research.

\bibliography{custom}

\appendix

\section{Trap–Audit–Recover Procedure}
\label{app:alg}

This section provides the full algorithmic description of the trap–audit–recover loop implemented by \tool{} (\Aref{alg:agentvisor}). It presents the end-to-end control flow, including proposal, auditing, exception handling, and final execution.

\begin{algorithm}[!t]
\caption{Trap--Audit--Recover for Tool Calls}
\label{alg:agentvisor}
\KwIn{Trusted system instruction $I_{\text{sys}}$, user query $I_u$, sanitized history $\tilde{H}_{t-1}$, Guest proposal $T_t^{\text{raw}}$}
\KwOut{Executed tool call $T_t$ (or $\emptyset$)}

$(\textsf{dec}_t, E_t) \leftarrow \textsf{STI\_Audit}(I_{\text{sys}}, I_u, \tilde{H}_{t-1}, T_t^{\text{raw}})$\;
\If{$\textsf{dec}_t=\textsf{allow}$}{
    Execute $T_t \leftarrow T_t^{\text{raw}}$\;
    \Return{$T_t$}\;
}
\Else{
    \tcp{Semantic exception injection}
    Provide $E_t$ to the Guest and request a revised call $T_t'$ satisfying constraints in $E_t$\;
    Execute $T_t \leftarrow T_t'$\;
    \Return{$T_t$}\;
}
\end{algorithm}

\section{Experimental Results on GLM-4.7 Backbone}
\label{app:glm_results}

In this section, we present the full experimental results for the agent powered by the GLM-4.7 backbone. The defense results against direct prompt injection attacks are presented in \Tref{tab:direct_glm}, and the defense results against indirect prompt injection attacks are shown in \Tref{tab:indirect_glm}.

\begin{table*}[!t]
\caption{Defense performance (\%) of our method and baseline approaches under \textbf{direct injection attack}, evaluated across the \texttt{No Attack} setting and 7 representative attack methods.}
\label{tab:direct_glm}
\resizebox{\textwidth}{!}{
\begin{tabular}{@{}c|c|cc|cc|cc|cc|cc|cc|cc@{}}
\toprule
Method           & No Attack & \multicolumn{2}{c|}{Direct} & \multicolumn{2}{c|}{Ignore} & \multicolumn{2}{c|}{Escape} & \multicolumn{2}{c|}{Fakecom} & \multicolumn{2}{c|}{Combined} & \multicolumn{2}{c|}{SYSTEM} & \multicolumn{2}{c}{Important} \\ \midrule
Metric           & BU        & UA           & ASR          & UA           & ASR          & UA           & ASR          & UA            & ASR          & UA            & ASR           & UA           & ASR          & UA            & ASR           \\ \midrule
No Defense       & 82.75   & 82.25      & 78.35      & 42.00      & 72.00      & 84.00      & 82.00      & 44.15       & 71.00      & 0.00        & 96.00       & 85.71      & 22.44      & 88.00       & 86.00       \\ \midrule
Sandwich         & 89.79   & 83.67      & 46.93      & 79.59      & 12.24      & 77.50      & 75.51      & 51.02       & 48.97      & 32.65       & 46.83       & 85.79      & 0.00       & 85.71       & 77.55       \\
Instructional    & 81.65   & 89.45      & 55.45      & 87.45      & 2.34       & 81.65      & 77.55      & 75.51       & 67.34      & 16.32       & 46.93       & 91.83      & 2.04       & 77.55       & 26.53       \\
Reminder         & 89.79   & 89.00      & 30.61      & 83.51      & 0.00       & 91.84      & 89.79      & 71.42       & 48.97      & 28.57       & 61.22       & 89.79      & 0.00       & 95.95       & 8.16        \\
Isolation        & 81.25   & 79.59      & 24.48      & 65.30      & 10.28      & 69.38      & 53.06      & 51.02       & 16.32      & 18.35       & 57.14       & 91.93      & 0.00       & 81.63       & 71.42       \\
Spotlight        & 65.30   & 24.00      & 0.00       & 10.20      & 0.00       & 20.15      & 0.00       & 20.95       & 0.00       & 6.11        & 8.16        & 40.80      & 0.00       & 10.01       & 0.00        \\
DeBERTa & 80.00   & 80.75      & 69.74      & 29.00      & 68.36      & 77.87      & 76.37      & 30.00       & 65.19      & 0.00        & 85.31       & 74.00      & 19.78      & 80.13       & 80.87       \\
DataSentinel     & 78.25   & 0.00       & 0.00       & 0.00       & 0.00       & 0.00       & 0.00       & 0.00        & 0.00       & 0.00        & 0.00        & 0.00       & 0.00       & 0.00        & 1.87        \\ \midrule
Ours             & 82.00   & 67.25      & 0.00       & 83.68      & 0.00       & 88.72      & 0.00       & 69.42       & 0.00       & 74.36       & 0.00        & 83.95      & 0.00       & 90.17       & 0.00        \\ \bottomrule
\end{tabular}
}
\end{table*}

\begin{table*}[!t]
\caption{Defense performance (\%) of our method and baseline approaches under \textbf{indirect injection attack}, evaluated across the \texttt{No Attack} setting and 7 representative attack methods.}
\label{tab:indirect_glm}
\resizebox{\textwidth}{!}{
\begin{tabular}{@{}c|c|cc|cc|cc|cc|cc|cc|cc@{}}
\toprule
Method           & No Attack & \multicolumn{2}{c|}{Direct} & \multicolumn{2}{c|}{Ignore} & \multicolumn{2}{c|}{Escape} & \multicolumn{2}{c|}{Fakecom} & \multicolumn{2}{c|}{Combined} & \multicolumn{2}{c|}{SYSTEM} & \multicolumn{2}{c}{Important} \\ \midrule
Metric           & BU        & UA            & ASR         & UA            & ASR         & UA            & ASR         & UA            & ASR          & UA             & ASR          & UA            & ASR         & UA            & ASR           \\ \midrule
No Defense       & 87.50   & 79.00       & 6.00      & 90.50       & 2.11      & 89.00       & 5.25      & 90.67       & 3.92       & 89.67        & 1.03       & 82.13       & 3.11      & 88.67       & 15.18       \\ \midrule
Sandwich         & 87.50   & 79.00       & 3.00      & 87.50       & 1.75      & 85.00       & 4.00      & 88.00       & 2.45       & 81.75        & 0.89       & 90.00       & 1.50      & 90.00       & 11.25       \\
Instructional    & 87.50   & 78.25       & 6.00      & 83.85       & 2.00      & 87.25       & 5.05      & 81.05       & 3.06       & 92.75        & 1.00       & 90.75       & 2.00      & 90.75       & 22.97       \\
Reminder         & 85.00   & 79.00       & 1.05      & 87.05       & 1.38      & 90.05       & 2.85      & 91.00       & 1.35       & 92.25        & 1.15       & 91.25       & 1.05      & 90.00       & 6.42        \\
Isolation        & 80.00   & 90.75       & 2.08      & 90.50       & 1.56      & 89.74       & 1.45      & 90.15       & 1.85       & 90.17        & 0.50       & 89.75       & 1.35      & 92.25       & 11.03       \\
Spotlight        & 85.00   & 78.25       & 0.00      & 80.00       & 1.05      & 79.25       & 0.00      & 80.00       & 0.00       & 77.00        & 1.00       & 70.50       & 1.00      & 47.72       & 0.50        \\
DeBERTa & 50.00   & 41.00       & 0.00      & 43.00       & 0.00      & 44.00       & 1.00      & 44.00       & 1.00       & 41.00        & 0.11       & 42.00       & 0.00      & 28.00       & 3.56        \\
MELON            & 87.50   & 40.00       & 0.00      & 70.00       & 1.00      & 59.50       & 0.45      & 59.50       & 0.56       & 74.12        & 1.13       & 63.29       & 1.00      & 53.25       & 0.00        \\ \midrule
Ours             & 90.00   & 91.00       & 1.55      & 91.00       & 0.00      & 89.67       & 0.78      & 90.67       & 0.11       & 90.17        & 0.00       & 85.50       & 1.00      & 87.07       & 0.91        \\ \bottomrule
\end{tabular}
}
\end{table*}

\end{document}